\documentclass[11pt,a4paper]{article}
\pdfoutput=1
\usepackage[utf8]{inputenc}
\usepackage{jheppub}
\hypersetup{unicode=true,bookmarksopen=true}

\usepackage{amsthm}
\usepackage{mathtools}
\usepackage{lmodern}
\usepackage[T1]{fontenc}
\usepackage{bbm}
\DeclareMathAlphabet{\mathbfi}{OML}{cmm}{b}{it}
\usepackage[scr=rsfso]{mathalpha}

\let\originalleft\left
\let\originalright\right
\renewcommand{\left}{\mathopen{}\mathclose\bgroup\originalleft}
\renewcommand{\right}{\aftergroup\egroup\originalright}
\makeatletter
\newcommand{\biggg}{\bBigg@\thr@@}
\newcommand{\Biggg}{\bBigg@{3.5}}
\makeatother

\makeatletter
\newenvironment{equations}[1][]{\subequations\ifx\relax#1\relax\else\label{#1}\fi\align\ignorespaces}{\endalign\ignorespacesafterend\endsubequations}
\def\@spliteq#1{\begin{equation}\begin{split}#1\end{split}\end{equation}}
\def\@spliteqstar#1{\begin{equation*}\begin{split}#1\end{split}\end{equation*}}
\def\splitequation{\collect@body\@spliteq}
\expandafter\def\csname splitequation*\endcsname{\collect@body\@spliteqstar}

\expandafter\def\csname endsplitequation*\endcsname{\ignorespacesafterend}
\makeatother

\renewcommand{\vec}[1]{{\ifnum9<1#1\mathbf{#1}\else\ifcat\noexpand#1\relax\boldsymbol{#1}\else\mathbfi{#1}\fi\fi}}
\newcommand{\mathe}{\mathrm{e}}
\newcommand{\mathi}{\mathrm{i}}
\let\oldre\Re
\let\oldim\Im
\renewcommand{\Re}{\oldre\mathfrak{e}\,}
\renewcommand{\Im}{\oldim\mathfrak{m}\,}
\newcommand{\total}{\mathop{}\!\mathrm{d}}

\newcommand{\abs}[1]{{\left\lvert{#1}\right\rvert}}
\newcommand{\norm}[1]{{\left\lVert{#1}\right\rVert}}
\newcommand{\sgn}{\operatorname{sgn}}

\newcommand{\1}{\mathbbm{1}}
\newcommand{\tr}{\operatorname{tr}}

\newcommand{\eqend}[1]{\,#1}

\makeatletter
\def\bra{\@ifnextchar[\bra@size\bra@nosize}
\def\bra@size[#1]#2{#1\langle{#2}#1\rvert}
\def\bra@nosize#1{\left\langle{#1}\right\rvert}
\def\ket{\@ifnextchar[\ket@size\ket@nosize}
\def\ket@size[#1]#2{#1\lvert{#2}#1\rangle}
\def\ket@nosize#1{\left\lvert{#1}\right\rangle}
\makeatother

\newcommand{\supp}{\operatorname{supp}}

\allowdisplaybreaks

\makeatletter
\gdef\@fpheader{\strut}
\makeatother

\newtheorem{definition}{Definition}
\newtheorem{prop}{Proposition}

\begin{document}

\title{Petz--Rényi relative entropy in QFT from modular theory}

\author{Markus B. Fr{\"o}b}
\author{and Leonardo Sangaletti}

\affiliation{Institut f{\"u}r Theoretische Physik, Universit{\"a}t Leipzig, Br{\"u}derstra{\ss}e 16, 04103 Leipzig, Germany}

\emailAdd{mfroeb@itp.uni-leipzig.de}
\emailAdd{leonardo.sangaletti@uni-leipzig.de}

\abstract{We consider the generalization of the Araki--Uhlmann formula for relative entropy to Petz--Rényi relative entropy. We compute this entropy for a free scalar field in the Minkowski wedge between the vacuum and a coherent state, as well as for the free chiral current in a thermal state. In contrast to the relative entropy which in these cases only depends on the sympletic form and thus reduces to the classical entropy of a wave packet, the Petz--Rényi relative entropy also depends on the symmetric part of the two-point function and is thus genuinely quantum. We also consider the relation with standard subspaces, where we define the Rényi entropy of a vector and show that it admits an upper bound given by the entropy of the vector.}


\maketitle

\section{Introduction}
\label{sec:intro}

Using Tomita--Takesaki modular theory~\cite{tomita1967,takesaki1970}, Araki and Uhlmann~\cite{araki1975,araki1976,uhlmann1977} have shown that the relative entropy between two cyclic and separating states $\Psi$ and $\Phi$ can be computed as the expectation value of the relative modular Hamiltonian $\ln \Delta_{\Psi\vert\Phi}$ according to
\begin{equation}
\label{eq:araki_uhlmann}
\mathcal{S}(\Psi\Vert\Phi) = - \bra{\Psi} \ln \Delta_{\Psi\vert\Phi} \ket{\Psi} \eqend{.}
\end{equation}
The relative modular Hamiltonian depends on the two states as well as the von Neumann algebra $\mathfrak{A}$ describing the part of the system that one is interested in. In applications, $\mathfrak{A}$ is usually the algebra of fields restricted to a certain spacetime region, in which case the above formula computes the relative entanglement entropy between this region and its complement. Using~\eqref{eq:araki_uhlmann}, relative entropy has been computed in a number of examples~\cite{casinigrillopontello2019,longo2019,froebmuchpapadopoulos2023,dangeloetal2023}. Relative entropy and more generally the modular Hamiltonian have also been useful in deriving various constraints on quantum field theory in diverse settings. The literature on this topic is vast, and we refer the reader to the recent works~\cite{casinitestetorroba2017,hollandsishibashi2019,dowlingfloerchingerhaas2020,dangelo2021,morinellitanimotowegener2022,casinihuerta2023,casinisalazartorroba2023,schroeflfloerchinger2023,katsinispastras2024,abatetorroba2024,longomorinelli2024} for various aspects and references to earlier work.

Formula~\eqref{eq:araki_uhlmann} is a direct generalization of the quantum-mechanical one, where the relative entropy between two density matrices $\rho$ and $\sigma$ is given by
\begin{equation}
\label{eq:relative_entropy}
\mathcal{S}(\rho\Vert\sigma) = \tr\left( \rho \ln \rho - \rho \ln \sigma \right) \eqend{.}
\end{equation}
Namely, on the tensor product Hilbert space that describes the bipartite quantum system (of the part of interest and its complement) the relative modular Hamiltonian has the very simple form~\cite{witten2018}
\begin{equation}
\label{eq:modular_hamiltonian_qm}
\ln \Delta_{\Psi\vert\Phi} = \ln\left( \rho_\Phi \otimes \rho_\Psi^{-1} \right) \eqend{,}
\end{equation}
where $\rho_\Phi$ and $\rho_\Psi$ are the reduced density matrices obtained from the pure states $\Phi$ and $\Psi$ defined on the tensor product space by tracing over either the part of interest or its complement. It is then easy to see that expression~\eqref{eq:araki_uhlmann} reduces to~\eqref{eq:relative_entropy}, and in fact this was the original motivation for~\eqref{eq:araki_uhlmann}.

Operationally, relative entropy measures the difference between the two states in the sense that the probability to wrongly ascribe the state $\Phi$ to the system in the state $\Psi$ after $N$ measurements decays asymptotically for large $N$ like $\mathe^{- N \mathcal{S}(\Psi\Vert\Phi)}$~\cite{hiaipetz1991}. However, there are other measures of the difference between states, such as the Petz--Rényi relative entropy~\cite{renyi1961,petz1985,petz1986}
\begin{equation}
\label{eq:renyi_entropy}
\mathcal{S}_\alpha(\rho\Vert\sigma) = \frac{1}{\alpha-1} \ln \tr\left( \rho^\alpha \sigma^{1-\alpha} \right) \eqend{,}
\end{equation}
defined for $\alpha \in (0,1)$.\footnote{If $\rho \leq C \sigma$ for some constant $C > 0$, the definition can be further extended to $\alpha \in (1,\infty)$.} In the limit $\alpha \to 1$, one recovers the relative entropy~\eqref{eq:relative_entropy}. Since $\rho$ and $\sigma$ do not commute, other generalizations are possible which in a limit reduce to the relative entropy, such as sandwiched Rényi relative entropy~\cite{muellerlennertetal2013,wildewinteryang2014}
\begin{equation}
\label{eq:sandwiched_renyi_entropy}
\tilde{\mathcal{S}}_\alpha(\rho\Vert\sigma) = \frac{1}{\alpha-1} \ln \tr\left( \sigma^\frac{1-\alpha}{2\alpha} \rho \sigma^\frac{1-\alpha}{2\alpha} \right)^\alpha
\end{equation}
defined for $\alpha \in (0,1) \cup (1,\infty)$. Using the formula~\eqref{eq:modular_hamiltonian_qm}, it is easy to see that the Petz--Rényi relative entropy~\eqref{eq:renyi_entropy} can be written as~\cite{petz1985,petz1986}
\begin{equation}
\label{eq:renyi_entropy_lndelta_pre}
\mathcal{S}_\alpha(\Psi\Vert\Phi) = \frac{1}{\alpha-1} \ln \bra{\Psi} \Delta_{\Psi\vert\Phi}^{1-\alpha} \ket{\Psi} \eqend{,}
\end{equation}
while for the sandwiched Rényi relative entropy~\eqref{eq:sandwiched_renyi_entropy} such a generalization is much harder~\cite{jencova2018,bertascholztomamichel2018,casinimedinasalazartorroba2018,hiai2019,kato2024}. Let us note that this equation is to be understood from spectral calculus, which means that we actually define
\begin{equation}
\label{eq:renyi_entropy_lndelta}
\mathcal{S}_\alpha(\Psi\Vert\Phi) = \frac{1}{\alpha-1} \ln \int_0^\infty \lambda^{1-\alpha} \total \bra{\Psi} E_\lambda \ket{\Psi} \eqend{,}
\end{equation}
where $E_\lambda$ is the spectral resolution of the (positive) operator $\Delta_{\Psi\vert\Phi}$. In the remainder of this work, we thus concentrate on the Petz--Rényi relative entropy in the form~\eqref{eq:renyi_entropy_lndelta}. We show that it is well-defined in general for $\alpha \in [0,1)$ and that the limit $\alpha \to 1^-$ exists, that it can be computed using analytic continuation of the modular flow, and that it can be obtained for free fields using the standard subspace approach. Lastly, we consider the concrete example of a free scalar field in the Minkowski wedge.

\section{Petz--Rényi relative entropy}
\label{sec:Petz_entropy}

\subsection{General results}
\label{sec:Petz_entropy_general}

We recall that the relative Tomita operator $S_{\Psi\vert\Phi}$ is defined as the closure of the map
\begin{equation}
\label{eq:relative_tomita}
S_{\Psi\vert\Phi} a \ket{\Psi} = a^\dagger \ket{\Phi} \quad\text{for all}\quad a \in \mathfrak{A}
\end{equation}
for a von Neumann algebra $\mathfrak{A}$ acting on a Hilbert space $\mathcal{H}$ and two states $\ket{\Psi}, \ket{\Phi} \in \mathcal{H}$, which we both assume for simplicity to be cyclic and separating for $\mathfrak{A}$ as well as normalized. The polar decomposition $S_{\Psi\vert\Phi} = J_{\Psi\vert\Phi} \Delta^\frac{1}{2}_{\Psi\vert\Phi}$ then defines the relative modular conjugation $J_{\Psi\vert\Phi}$ and the relative modular operator $\Delta_{\Psi\vert\Phi}$. It also follows that $\ket{\Psi} \in \mathcal{D}\left( \Delta^\frac{1}{2}_{\Psi\vert\Phi} \right)$, which we will use in the following without explicitly mentioning it.

Since $\Delta_{\Psi\vert\Phi}$ is positive and $\lambda^r < 1 + \lambda$ for $r \in [0,1]$ and $\lambda \geq 0$, we obtain
\begin{splitequation}
\label{eq:estimate_bound}
0 &\leq \int_0^\infty \lambda^r \total \bra{\Psi} E_\lambda \ket{\Psi} < \int_0^\infty (1 + \lambda) \total \bra{\Psi} E_\lambda \ket{\Psi} \\
&= \norm{ \ket{\Psi} }^2 + \norm{ \Delta_{\Psi\vert\Phi}^\frac{1}{2} \ket{\Psi} }^2 = 1 + \norm{ J_{\Psi\vert\Phi}^{-1} \ket{\Phi} }^2 = 1 + \norm{ \ket{\Phi} }^2 = 2 \eqend{.}
\end{splitequation}
We see that the spectral integral is uniformly bounded (independently of $r$), and it follows that it is continuous as a function of $r \in [0,1]$, keeping in mind that the states $\ket{\Psi}$ and $\ket{\Phi}$ are fixed. We would also like to take derivatives with respect to $r$, which we may exchange with the integration by dominated convergence provided we can find a dominating integrable function. For this, we note that for $k \in \mathbb{N}$ we have
\begin{splitequation}
\abs{ \lambda^r \ln^k \lambda } &\leq \Theta(1-\lambda) \lambda^r \ln^k \lambda^{-1} + \Theta(\lambda - 1) \lambda^r \ln^k \lambda \\
&\leq \Theta(1-\lambda) \ln^k \lambda^{-1} + \Theta(\lambda - 1) \frac{\lambda^{r+k\epsilon}}{\epsilon^k \, \mathe^k}
\end{splitequation}
for all $\epsilon > 0$. This is an integrable function for all $r \in [0,r_1]$ with $0 < r_1 < 1$, where we may take $\epsilon = (1 - r_1)/k$. Since $r_1$ is arbitrary, this shows that arbitrary derivatives with respect to $r$ can be taken inside the integration in the range $r \in [0,1)$:
\begin{equation}
\partial_r^k \int_0^\infty \lambda^r \total \bra{\Psi} E_\lambda \ket{\Psi} = \int_0^\infty \lambda^r \ln^k \lambda \total \bra{\Psi} E_\lambda \ket{\Psi} \eqend{.}
\end{equation}
However, continuity of the derivatives as $r \to 1$, which corresponds to $\alpha \to 0$, is not guaranteed.

It follows that the Petz--Rényi relative entropy $\mathcal{S}_\alpha(\Psi\Vert\Phi)$~\eqref{eq:renyi_entropy_lndelta} is well-defined for $\alpha \in [0,1)$. In the limit $\alpha \to 1^-$, l'H{\^o}pital's rule shows that
\begin{splitequation}
\label{eq:renyi_entropy_limit}
\lim_{\alpha \to 1^-} \mathcal{S}_\alpha(\Psi\Vert\Phi) &= \lim_{\alpha \to 1^-} \partial_\alpha \ln \int_0^\infty \lambda^{1-\alpha} \total \bra{\Psi} E_\lambda \ket{\Psi} \\
&= - \lim_{\alpha \to 1^-} \left[ \left( \int_0^\infty \lambda^{1-\alpha} \total \bra{\Psi} E_\lambda \ket{\Psi} \right)^{-1} \int_0^\infty \lambda^{1-\alpha} \ln \lambda \total \bra{\Psi} E_\lambda \ket{\Psi} \right] \\
&= - \int_0^\infty \ln \lambda \total \bra{\Psi} E_\lambda \ket{\Psi} = - \bra{\Psi} \ln \Delta_{\Psi\vert\Phi} \ket{\Psi} = \mathcal{S}(\Psi\Vert\Phi) \eqend{,}
\end{splitequation}
where we also used that $\int_0^\infty \total \bra{\Psi} E_\lambda \ket{\Psi} = \norm{ \ket{\Psi} }^2 = 1$. That is, in the limit $\alpha \to 1$ the Petz--Rényi relative entropy reduces to the Araki--Uhlmann relative entropy~\eqref{eq:araki_uhlmann}, which is known to be positive. Note that for $\alpha = 0$ we have
\begin{splitequation}
\mathcal{S}_0(\Psi\Vert\Phi) &= - \ln \int_0^\infty \lambda \total \bra{\Psi} E_\lambda \ket{\Psi} = - \ln \norm{ \Delta_{\Psi\vert\Phi}^\frac{1}{2} \ket{\Psi} }^2 \\
&= - \ln \norm{ J_{\Psi\vert\Phi} S_{\Psi\vert\Phi} \ket{\Psi} }^2 = - \ln \norm{ J_{\Psi\vert\Phi} \ket{\Phi} }^2 = - \ln \norm{ \ket{\Phi} }^2 = 0 \eqend{.}
\end{splitequation}
To show that Petz--Rényi relative entropy $\mathcal{S}_\alpha(\Psi\Vert\Phi)$ is non-negative for $\alpha \in [0,1]$, we can use that it is monotonically increasing with $\alpha$~\cite[App.~B]{bertascholztomamichel2018}. This follows because $\frac{F(r)}{r}$ is monotonically decreasing for $r \in (0,1)$ if $F(0) \geq 0$ and $F$ is a concave function, since then
\begin{equation}
F(a) = F\left( \frac{a}{b} b + \left( 1 - \frac{a}{b} \right) 0 \right) \geq \frac{a}{b} F(b) + \left( 1 - \frac{a}{b} \right) F(0) \geq \frac{a}{b} F(b) \quad\text{for}\quad 0 \leq a \leq b \eqend{.}
\end{equation}
So it remains to show that $(1-\alpha) \mathcal{S}_\alpha(\Psi\Vert\Phi)$ is concave as a function of $r = 1-\alpha$, i.e., that
\begin{equation}
F(r) = - \ln \int_0^\infty \lambda^r \total \bra{\Psi} E_\lambda \ket{\Psi}
\end{equation}
is concave, and that $F(0) \geq 0$. The second property follows easily from
\begin{equation}
F(0) = - \ln \int_0^\infty \total \bra{\Psi} E_\lambda \ket{\Psi} = - \ln \norm{ \ket{\Psi} }^2 = 0 \eqend{.}
\end{equation}
For concavity, we use that since $F(r)$ is a differentiable function for $r \in (0,1)$ by the above results, we can compute its second derivative which reads
\begin{splitequation}
&\left[ \int_0^\infty \lambda^r \total \bra{\Psi} E_\lambda \ket{\Psi} \right]^2 F''(r) \\
&= \left[ \left( \int_0^\infty \lambda^r \ln \lambda \total \bra{\Psi} E_\lambda \ket{\Psi} \right)^2 - \int_0^\infty \lambda^r \total \bra{\Psi} E_\lambda \ket{\Psi} \int_0^\infty \lambda^r \ln^2 \lambda \total \bra{\Psi} E_\lambda \ket{\Psi} \right] \\
&= - \frac{1}{2} \int_0^\infty \int_0^\infty \lambda^r \mu^r \left( \ln \lambda - \ln \mu \right)^2 \total \bra{\Psi} E_\lambda \ket{\Psi} \total \bra{\Psi} E_\mu \ket{\Psi} \leq 0 \eqend{,}
\end{splitequation}
and it follows that $F(r)$ is concave as required.

Lastly, we would like to show that one can compute the Petz--Rényi entropy using an analytic continuation of the modular flow. Consider thus the function
\begin{equation}
\label{eq:modular_flow_f}
f(t) = \bra{\Psi} \Delta_{\Psi\vert\Phi}^{\mathi t} \ket{\Psi} = \int_0^\infty \lambda^{\mathi t} \total \bra{\Psi} E_\lambda \ket{\Psi} \eqend{,}
\end{equation}
defined initially for $t \in \mathbb{R}$. We want to prove that it admits an analytic extension into the complex strip
\begin{equation}
\label{eq:complex_strip}
\mathscr{I} = \{ z = t - \mathi r \in \mathbb{C}\colon t \in \mathbb{R}, r \in (0,1) \}
\end{equation}
with a continuous extension to the boundaries $r = 0$ and $r = 1$. The proof of the existence of such an extension substantially relies on the the analyticity of the exponential function, and we follow~\cite[Lemma~9.2.12]{kadison1997fundamentals}. First of all, we notice that the function $f(t)$ can be extended to a bounded function of $z$ on the strip $\mathscr{I}$, thanks to the bound $\abs{\lambda^{\mathi z}} = \lambda^r$ and the estimates~\eqref{eq:estimate_bound} for $r \in [0,1]$. To show that this extension is also analytic, we define a family of complex-valued functions $\{ f_n \}$ for $n \in \mathbb{N}$ by
\begin{equation}
f_n(z) = \int_{[n^{-1},n]} \lambda^{\mathi z} \total \bra{\Psi} E_\lambda \ket{\Psi} \eqend{.}
\end{equation}
For every $n \in \mathbb{N}$ it is easy to see that the functions $f_n(z)$ are analytic on $\mathbb{C}$ thanks to the uniform bound $\abs{ \partial_z^k f_n(z) } \leq n^{\abs{\Im z}} \ln^k n$. In addition, for $z \in \mathscr{I}$ it holds that
\begin{equation}
\abs{ f(z) - f_n(z) } \leq \int_{[0,n^{-1}) \cup (n,\infty)} (1+\lambda) \total \bra{\Psi} E_\lambda \ket{\Psi} \leq \int_0^\infty (1+\lambda) \total \bra{\Psi} E_\lambda \ket{\Psi} = 2 \eqend{,}
\end{equation}
using that $\abs{ \lambda^{\mathi z} } \leq \lambda^r < 1 + \lambda$. The difference between $f(z)$ and $f_n(z)$ is thus uniformly bounded, and we can apply the dominated convergence theorem to conclude that the sequence $\{ f_n \}$ converges uniformly to $f(z)$ for $z \in \mathscr{I}$. It then follows that the extension $f(z)$ is analytic in the strip $\mathscr{I}$ and continuous on the boundaries.

Summarizing the above results, we obtain
\begin{prop}
\label{prop:entropy_properties}
The Petz--Rényi relative entropy $\mathcal{S}_\alpha(\Psi\Vert\Phi)$ defined by~\eqref{eq:renyi_entropy_lndelta} is well-defined for $\alpha \in [0,1]$, and a differentiable function of $\alpha$ for all $\alpha \in [0,1)$. It vanishes for $\alpha = 0$ and increases monotonically with $\alpha$, and is thus non-negative for all $\alpha \in [0,1]$ (but may become infinite). In the limit $\alpha \to 1^-$, it reduces to the Araki--Uhlmann relative entropy~\eqref{eq:araki_uhlmann}, which is thus an upper bound for the Petz--Rényi relative entropy. It can be computed by analytically continuing the modular flow~\eqref{eq:modular_flow_f}.
\end{prop}

In general, relative modular operators and Hamiltonians are difficult to compute explicitly. One exception is if both $\ket{\Psi}$ and $\ket{\Phi}$ are excitations of a common state $\ket{\Omega}$, which is also cyclic and separating for $\mathfrak{A}$. That is, assuming that $\ket{\Psi} = U U' \ket{\Omega}$ and $\ket{\Phi} = V V' \ket{\Omega}$ with invertible $U,V \in \mathfrak{A}$ and invertible $U',V' \in \mathfrak{A}'$, equation~\eqref{eq:relative_tomita} for the relative Tomita operator can be written as
\begin{splitequation}
U^\dagger (V')^{-1} S_{\Psi\vert\Phi} U' (V^{-1})^\dagger V^\dagger a U \ket{\Omega} &= U^\dagger (V')^{-1} S_{\Psi\vert\Phi} a U U' \ket{\Omega} = U^\dagger (V')^{-1} S_{\Psi\vert\Phi} a \ket{\Psi} \\
&= U^\dagger (V')^{-1} a^\dagger \ket{\Phi} = U^\dagger (V')^{-1} a^\dagger V V' \ket{\Omega} \\
&= (V^\dagger a U)^\dagger \ket{\Omega} = S_\Omega V^\dagger a U \ket{\Omega} \eqend{,}
\end{splitequation}
where we used that $U'$ and $V'$ are in the commutant $\mathfrak{A}'$ and thus commute with $a,U,V$ and their adjoints which are in the algebra $\mathfrak{A}$. Since $U$ and $V$ are invertible, this is equivalent to
\begin{equation}
\left( U^\dagger (V')^{-1} S_{\Psi\vert\Phi} U' (V^{-1})^\dagger - S_\Omega \right) b \ket{\Omega} = 0 \quad\text{for all}\quad b \in \mathfrak{A} \eqend{,}
\end{equation}
and since $\ket{\Omega}$ is cyclic for $\mathfrak{A}$ such that the set $\{ b \ket{\Omega}\colon b \in \mathfrak{A} \}$ is dense in $\mathcal{H}$, we obtain
\begin{equation}
S_{\Psi\vert\Phi} = (U^{-1})^\dagger V' S_\Omega (U')^{-1} V^\dagger \eqend{.}
\end{equation}
The relative modular Hamiltonian and relative modular conjugation can then be obtained from the polar decomposition of $S_{\Psi\vert\Phi}$. In particular if $U$ and $V'$ are unitary, it follows that
\begin{equation}
\label{eq:relative_modular_delta}
\Delta_{\Psi\vert\Phi} = S_{\Psi\vert\Phi}^\dagger S_{\Psi\vert\Phi} = [ (U')^{-1} V^\dagger ]^\dagger \Delta_\Omega (U')^{-1} V^\dagger \eqend{,}
\end{equation}
and if also $U'$ and $V$ are unitary, spectral calculus shows that
\begin{equation}
\label{eq:relative_modular_delta_alpha}
\Delta_{\Psi\vert\Phi}^r = V U' \Delta_\Omega^r (U')^\dagger V^\dagger \eqend{.}
\end{equation}
For unitary excitations of a common state $\ket{\Omega}$, the Petz--Rényi relative entropy~\eqref{eq:renyi_entropy_lndelta} therefore reduces to
\begin{equation}
\label{eq:renyi_entropy_lndelta_unitary}
\mathcal{S}_\alpha(U U' \Omega\Vert V V' \Omega) = \frac{1}{\alpha-1} \ln \int_0^\infty \lambda^{1-\alpha} \total \bra{V^\dagger U \Omega} E_\lambda \ket{V^\dagger U \Omega} \eqend{,}
\end{equation}
where now $E_\lambda$ is the spectral resolution of the (positive) operator $\Delta_\Omega$. In particular, it does not depend on $U'$ and $V'$, the excitations in the commutant. This shows that the Petz--Rényi relative entropy, similarly to the Araki--Uhlmann relative entropy, only depends on the states defined on the algebra $\mathfrak{A}$ and not on their specific vector representatives in the Hilbert space $\mathcal{H}$. Indeed, the two vectors $\ket{\Omega}$ and $U' \ket{\Omega}$ define the same state $\omega$ on $\mathfrak{A}$ via $\omega(a) = \bra{\Omega} a \ket{\Omega} = \bra{\Omega} (U')^\dagger a U' \ket{\Omega}$ for all $a \in \mathfrak{A}$.

\subsection{Free bosonic QFTs}

Let us now specialize to a free bosonic quantum field theory, and the corresponding second-quantized symmetric Fock space $\mathcal{F}$ constructed over the one-particle Hilbert space $\mathcal{H}$. On $\mathcal{F}$ there exists a representation of the abstract Weyl algebra, which is the completion of the free algebra generated by the identity $\1$ and the Weyl operators $W(f)$, quotiened by the relations
\begin{equation}
[ W(f) ]^* = W(-f) \eqend{,} \quad W(f) W(g) = \mathe^{- \frac{\mathi}{2} \sigma(f,g)} W(f+g) \eqend{.}
\end{equation}
Here, $f$ are elements of a symplectic space $(S,\sigma)$ and $\sigma$ is the corresponding symplectic form; in the case that we are interested in, $f$ are vectors in some real-linear subspace $\mathcal{L}$ of the one-particle Hilbert space $\mathcal{H}$ and $\sigma$ is twice the imaginary part of the scalar product $(\cdot,\cdot)_\mathcal{H}$ on $\mathcal{H}$ and non-degenerate. If the Hilbert space is not given, it is possible to construct it (and the symmetric Fock space) starting from the symplectic space together with a quasi-free state $\omega$ on the Weyl algebra $\mathcal{A}$~\cite[App.~A]{kaywald1991}. This state is defined by the action
\begin{equation}
\omega( W(f) ) = \mathe^{- \frac{1}{2} \mu(f,f)} \eqend{,}
\end{equation}
where $\mu$ is a real symmetric bilinear form on $S \times S$ that satisfies $\mu(f,f) \mu(h,h) \geq \frac{1}{4} \sigma(f,h)^2$ for all $f,h \in S$ to ensure positivity. From $\mu$ and $\sigma$, one defines the bilinear form $\omega_2 = \mu + \frac{\mathi}{2} \sigma$, which becomes the scalar product on the one-particle Hilbert space $\mathcal{H}$. Furthermore, the representation $\pi$ is regular such that $\pi(W(f)) = \mathe^{\mathi \phi(f)}$ with a self-adjoint operator $\phi(f)$ defined on a dense subset of $\mathcal{F}$~\cite[Sec.~5.2.3]{brattelirobinson2}, the state is represented by the cyclic vector $\ket{\Omega}$ through the relation
\begin{equation}
\omega( W(f) ) = \bra{\Omega} \pi(W(f)) \ket{\Omega} = \bra{\Omega} \mathe^{\mathi \phi(f)} \ket{\Omega} \eqend{,}
\end{equation}
and the von Neumann algebra $\mathfrak{A}$ is obtained as the weak closure $\mathfrak{A} = \pi(\mathcal{A})''$. If in addition the state $\omega$ is faithful for $\mathcal{A}$, the vector $\ket{\Omega}$ is also separating for $\mathfrak{A}$, and we will assume this in the following. Therefore, the modular objects can be constructed as before. Since in the following we will only work with this representation, for ease of notation we do not write $\pi$ explicitly.

At this point, we can derive a more explicit expression for the Petz--Rényi relative entropy between the state $\omega$ and a state $\psi$ obtained as a coherent excitation of $\omega$. Indeed, the state $\psi$ is implemented by a vector $\ket{\Psi} = W(f) \ket{\Omega}$, and since $W(f)$ is unitary we can employ equation~\eqref{eq:renyi_entropy_lndelta_unitary} with $U' = V' = V = \1$ and $U = W(f)$. Therefore we have
\begin{equation}
\mathcal{S}_\alpha( W(f) \Omega\Vert \Omega) = \frac{1}{\alpha-1} \ln \int_0^\infty \lambda^{1-\alpha} \total \bra{\Omega} W(-f) E_\lambda W(f) \ket{\Omega} \eqend{,}
\end{equation}
and as explained before we will compute this by analytically continuing the result for the modular flow. For free theories, it has been shown~\cite{figlioliniguido1989,figlioliniguido1994} that the modular objects are second-quantized operators on Fock space, which in particular means that the modular group acts on the Weyl generators according to
\begin{equation}
\label{eq:geometric_modular_action}
\Delta_\Omega^{\mathi t} W(f) \Delta_\Omega^{-\mathi t} = W(f_t) \quad\text{with}\quad f_t \in \mathcal{L} \ \forall t \in \mathbb{R} \eqend{.}
\end{equation}
This action holds even more generally, namely when the modular group acts in a local geometric way~\cite{buchholzsummers1993,buchholzdreyerflorigsummers2000} (for example, in the cases covered by the Bisognano--Wichmann theorem). We then obtain
\begin{splitequation}
\bra{\Omega} W(-f) \Delta_\Omega^{\mathi t} W(f) \ket{\Omega} &= \bra{\Omega} \Delta_\Omega^\frac{-\mathi t}{2} W(-f) \Delta_\Omega^\frac{\mathi t}{2} \Delta_\Omega^\frac{\mathi t}{2} W(f) \Delta_\Omega^\frac{-\mathi t}{2} \ket{\Omega} \\
&= \bra{\Omega} W\left( - f_{-\frac{t}{2}} \right) W\left( f_\frac{t}{2} \right) \ket{\Omega} = \bra{\Omega} \mathe^{\frac{\mathi}{2} \sigma\left( f_{-\frac{t}{2}}, f_\frac{t}{2} \right)} W\left( - f_{-\frac{t}{2}} + f_\frac{t}{2} \right) \ket{\Omega} \\
&= \mathe^{- \frac{1}{2} \omega_2\left( - f_{-\frac{t}{2}} + f_\frac{t}{2}, - f_{-\frac{t}{2}} + f_\frac{t}{2} \right)} \mathe^{\frac{\mathi}{2} \sigma\left( f_{-\frac{t}{2}}, f_\frac{t}{2} \right)} \eqend{,}
\end{splitequation}
where we used that the action of $\Delta_\Omega$ leaves the state $\ket{\Omega}$ invariant. Performing the analytic continuation and taking the logarithm, we obtain that the Petz--Rényi relative entropy is given by
\begin{equation}
\mathcal{S}_\alpha( W(f) \Omega \Vert \Omega ) = \frac{1}{\alpha-1} \ln M( \mathi (\alpha-1) ) \eqend{,}
\end{equation}
where $M$ is the function defined for real $t$ by
\begin{equation}
M(t) = \exp\left[ - \frac{1}{2} \omega_2\left( f_\frac{t}{2} - f_{-\frac{t}{2}}, f_\frac{t}{2} - f_{-\frac{t}{2}} \right) + \frac{\mathi}{2} \sigma\left( f_{-\frac{t}{2}}, f_\frac{t}{2} \right) \right] \eqend{,}
\end{equation}
and the analytic continuation is well-defined by our previous arguments. Using the relation $\mathi \sigma(f,g) = \omega_2(f,g) - \omega_2(g,f)$ between the antisymmetric part of the two-point function $\omega_2$ and the symplectic form as well as $\omega_2\left( f_t, f_t \right) = \omega_2\left( f, f \right)$, we can also express $M(t)$ as
\begin{equation}
M(t) = \exp\left[ \omega_2\left( f_{-\frac{t}{2}}, f_\frac{t}{2} \right) - \omega_2\left( f, f \right) \right] \eqend{.}
\end{equation}
Since we have shown that $\mathcal{S}_\alpha( W(f) \Omega \Vert \Omega ) \geq 0$, it follows that the analytic continuation of $M(t)$ to $M( \mathi (\alpha-1) )$ is a real function satisfying $0 \leq M( \mathi (\alpha-1) ) \leq 1$. Performing the analytic continuation through $\Re z = \Re (t-\mathi r) = t \geq 0$ and employing the principal branch of the logarithm, we can interchange the analytic continuation with taking the logarithm, and obtain
\begin{prop}
\label{prop:entropy_unitary_excitation}
Under the assumption~\eqref{eq:geometric_modular_action} of geometric modular action, the Petz--Rényi relative entropy between the vacuum $\ket{\Omega}$ and the unitary excitation $W(f) \ket{\Omega}$ is given by
\begin{equation}
\label{eq:renyi_relative_unitary}
\mathcal{S}_\alpha( W(f) \Omega \Vert \Omega ) = \frac{1}{\alpha-1} F( \mathi (\alpha-1) ) \eqend{,}
\end{equation}
where $F$ is the function defined for real $t$ by
\begin{splitequation}
\label{eq:renyi_relative_fdef}
F(t) &= - \frac{1}{2} \omega_2\left( f_\frac{t}{2} - f_{-\frac{t}{2}}, f_\frac{t}{2} - f_{-\frac{t}{2}} \right) + \frac{\mathi}{2} \sigma\left( f_{-\frac{t}{2}}, f_\frac{t}{2} \right) \\
&= \omega_2\left( f_{-\frac{t}{2}}, f_\frac{t}{2} \right) - \omega_2\left( f, f \right) \eqend{,}
\end{splitequation}
and the analytic continuation is well-defined.
\end{prop}
We see that in the limit $\alpha \to 1$, where the Petz--Rényi relative entropy reduces to the Araki--Uhlmann relative entropy, only the linear expansion of $F(t)$ around $t = 0$ is relevant. Because $\omega_2\left( f_\frac{t}{2} - f_{-\frac{t}{2}}, f_\frac{t}{2} - f_{-\frac{t}{2}} \right)$ vanishes to first order in $t$, from~\eqref{eq:renyi_relative_fdef}, we see that in this limit only the sympletic form $\sigma$ makes a contribution. Since the symplectic form already enters the classical analysis, while the symmetric part of the two-point function $\omega_2$ only becomes important in the quantum theory, in contrast to the Araki--Uhlmann relative entropy which can be reduced to the classical entropy of a wave packet~\cite{ciollilongoruzzi2020}, the Petz--Rényi relative entropy is a genuinely quantum measure of entropy.

\subsection{Relation to standard subspaces}

For free theories, both the modular data and relative entropy can be nicely determined using standard subspaces~\cite{longo2019,ciollilongoruzzi2020,bostelmanncadamurodelvecchio2022}. A standard subspace $\mathcal{L}$ of a complex Hilbert space $\mathcal{H}$ is a closed real-linear subspace such that $\mathcal{L} \cap I \mathcal{L} = \{ 0 \}$ and $\overline{\mathcal{L} + I \mathcal{L}} = \mathcal{H}$, where $I$ is the complex structure, a real anti-Hermitean operator satisfying $I^2 = - \1$. For such subspaces, one can define the modular objects, starting from the Tomita operator which is obtained as the closure of the map $s$ acting as $s (h + I k) = h - I k$ for $h,k \in \mathcal{L}$. The polar decomposition $s = j \delta^\frac{1}{2}$ of the (completion of) $s$ then defines the modular conjugation $j$ and the modular operator $\delta$. For simplicity we only treat the bosonic case and further assume that $\mathcal{L}$ is factorial, that is $\mathcal{L} \cap \mathcal{L}' = 0$ where $\mathcal{L}' = (I \mathcal{L})^{\perp_\mathbb{R}}$, with $\perp_\mathbb{R}$ denoting the orthogonal complement with respect to the real part of the scalar product on $\mathcal{H}$. It has been shown~\cite{figlioliniguido1989,figlioliniguido1994} that this condition is equivalent to the fact that $1$ is not in the point spectrum of the modular operator $\Delta$, and it will be fulfilled in our examples. Moreover, the Fock space modular operators $J_\Omega$ and $\Delta_\Omega$ for the Fock vacuum state $\ket{\Omega}$ are obtained as the second-quantized versions of $j$ and $\delta$ on the single-particle Hilbert space~\cite{figlioliniguido1989,figlioliniguido1994}. This means in particular that the modular group action on the Weyl generators~\eqref{eq:geometric_modular_action} on Fock space has an analogue in $\mathcal{H}$, namely for $f \in \mathcal{L}$ it holds that
\begin{equation}
\label{eq:modular_action_standardsubspace}
\delta^{\mathi t} f = f_t \quad\text{with}\quad f_t \in \mathcal{L} \ \forall t \in \mathbb{R} \eqend{,}
\end{equation}
and with the same $f_t$ that appears in equation~\eqref{eq:geometric_modular_action}.

With $f \in \mathcal{L}$, the Petz--Rényi relative entropy between the vacuum $\ket{\Omega}$ and the unitary excitation $W(f) \ket{\Omega}$ is given by Prop.~\ref{prop:entropy_unitary_excitation}. Using that the two-point function $\omega_2$, evaluated on functions $f \in \mathcal{H}$, is equal to the scalar product on the one-particle Hilbert space $\mathcal{H}$, we can rewrite the result~\eqref{eq:renyi_relative_fdef} as
\begin{equation}
F(t) = \left( f_{-\frac{t}{2}}, f_\frac{t}{2} \right)_\mathcal{H} - \left( f, f \right)_\mathcal{H} = \left( \delta^{- \frac{\mathi t}{2}} f, \delta^\frac{\mathi t}{2} f \right)_\mathcal{H} - \left( f, f \right)_\mathcal{H} \eqend{,}
\end{equation}
where we used the action~\eqref{eq:modular_action_standardsubspace} of the single-particle modular operator. The analytic continuation $t \to \mathi (\alpha-1)$ leads to
\begin{equation}
\mathcal{S}_\alpha( W(f) \Omega \Vert \Omega ) = \frac{1}{\alpha-1} \left[ \left( \delta^{- \frac{1-\alpha}{2}} f, \delta^\frac{1-\alpha}{2} f \right)_\mathcal{H} - \left( f, f \right)_\mathcal{H} \right] \eqend{.}
\end{equation}
Employing the spectral resolution $e_\lambda$ of $\delta$, we thus arrive at
\begin{definition}
\label{def:renyi_entropy_vector}
The Rényi entropy of a vector $f \in \mathcal{L}$ with respect to the standard subspace $\mathcal{L} \subset \mathcal{H}$ is defined as
\begin{equation}
\mathcal{S}_\alpha(f) = \int_0^\infty \frac{\lambda^{1-\alpha} - 1}{\alpha-1} \total \left( f, e_\lambda f \right)_\mathcal{H} \eqend{.}
\end{equation}
\end{definition}
Completely analogous to Prop.~\ref{prop:entropy_properties}, one shows that $\mathcal{S}_\alpha(f)$ is continuous in $\alpha$, and in the limit $\alpha \to 1^-$ it follows that
\begin{equation}
\mathcal{S}_\alpha(f) = - \int_0^\infty \ln \lambda \total \left( f, e_\lambda f \right)_\mathcal{H} \eqend{,}
\end{equation}
which is nothing else but the entropy of the vector $f \in \mathcal{L}$ as defined in~\cite{longo2019}. In analogy to Prop.~\ref{prop:entropy_properties}, one further shows that
\begin{prop}
\label{prop:entropy_properties_subspace}
The Rényi entropy of a vector $f \in \mathcal{L}$ is well-defined for $\alpha \in [0,1]$, and a differentiable function of $\alpha$ for all $\alpha \in [0,1)$. It vanishes for $\alpha = 0$ and increases monotonically with $\alpha$, and is thus non-negative for all $\alpha \in [0,1]$ (but may become infinite). In the limit $\alpha \to 1^-$, it reduces to the entropy of the vector $f \in \mathcal{L}$, which is thus an upper bound for the Rényi entropy of this vector.
\end{prop}

\section{Free scalar fields in a wedge}

We can now compute Petz--Rényi relative entropy for a free massive real scalar field in the Bisognano--Wichmann situation. More specifically, we consider the Minkowski vacuum representation (the GNS representation induced by the vacuum state $\omega$) of the scalar field. In this representation, we define the von Neumann algebra $\mathfrak{A}$ generated by the Weyl operators localized in the right wedge $\mathcal{W}_r = \{ x \in \mathbb{M}^{d+1}\colon x^1 > \abs{x^0} \}$, where $(\mathbb{M}^{d+1},\eta)$ is the $(d+1)$-dimensional Minkowski spacetime with metric $\eta$ (mostly plus in our conventions). Thanks to the Bisognano--Wichmann theorem~\cite{bisognanowichmann1975,bisognanowichmann1976}, we know that the vacuum vector $\ket{\Omega}$ is cyclic and separating for $\mathfrak{A}$ and the modular operator for the pair $(\mathfrak{A},\ket{\Omega})$ coincides with the boost operator in the positive $x^1$ direction. In particular, we have the geometric modular action~\eqref{eq:geometric_modular_action} with
\begin{equation}
f_t(x) = f(\Lambda_{-t} x) \eqend{,}
\end{equation}
where we denoted with $\Lambda_{-t}$ the boosted coordinates in the $x^1$ direction:
\begin{equations}[eq:boosted_coords]
(\Lambda_t x)^0 &= \cosh(2 \pi t) x^0 + \sinh(2 \pi t) x^1 \eqend{,} \\
(\Lambda_t x)^1 &= \sinh(2 \pi t) x^0 + \cosh(2 \pi t) x^1 \eqend{,} \\
(\Lambda_t x)^i &= x^i \quad\text{for}\quad i \in \{ 2,\ldots,d+1 \} \eqend{.}
\end{equations}
The integral kernel of the two-point function of the state $\omega$ is given by
\begin{equation}
\label{eq:omega_2pf}
\omega_2(x,y) = \lim_{\epsilon \to 0^+} \int \frac{1}{2 \omega_\vec{p}} \mathe^{\mathi \omega_\vec{p} (x^0-y^0 + \mathi \epsilon) - \mathi p^1 (x^1-y^1) - \mathi p^j (x^j-y^j)} \frac{\total^d \vec{p}}{(2 \pi)^d} \eqend{,}
\end{equation}
where $\omega_\vec{p} = \sqrt{ \vec{p}^2 + m^2 }$ and $j \in \{ 2,\ldots,d+1 \}$. To obtain an explicit expression for the Petz--Rényi relative entropy, we need to find the analytic extension of the function
\begin{equation}
t \in \mathbb{R} \to \omega_2\left( f_{-\frac{t}{2}}, f_{\frac{t}{2}} \right)
\end{equation}
in the complex strip $\mathscr{I}$. The existence of this analytic extension has been proven on a general ground in section~\ref{sec:Petz_entropy_general}. We first note that since any orthochronous Lorentz transformation has determinant one, we have
\begin{splitequation}
\omega_2(f_s, g_{s'}) &= \iint \omega_2(x,y) f(\Lambda_{-s} x) g(\Lambda_{-s'} y) \total x \total y \\
&= \iint \omega_2(\Lambda_s x, \Lambda_{s'} y) f(x) g(y) \total x \total y \eqend{,}
\end{splitequation}
which in particular entails that
\begin{equation}
\omega_2\left( f_{-\frac{t}{2}}, f_{\frac{t}{2}} \right) = \iint \omega_2\left( \Lambda_{-\frac{t}{2}} x, \Lambda_{\frac{t}{2}} y \right) f(x) g(y) \total x \total y \eqend{.}
\end{equation}
We can now extend the distributional kernel $\omega_2\left( \Lambda_{-\frac{t}{2}} x, \Lambda_{\frac{t}{2}} y \right)$ (as a function of the real variable $t$) to a bounded function in the complex plane in the following way: Let $z = a + \mathi b$ with $a,b \in \mathbb{R}$ and $w = c + \mathi d$ with $c,d \in \mathbb{R}$ be two complex numbers. By direct computation using~\eqref{eq:boosted_coords}, we obtain
\begin{equation}
\label{eq:boosted_2pf}
\omega_2(\Lambda_z x, \Lambda_w y) = \lim_{\epsilon \to 0^+} \int \frac{1}{2 \omega_\vec{p}} \exp\left[ \mathi H(x,y,z,w,p) - \epsilon \, \omega_\vec{p} \right] \, \mathe^{- \mathi p^j (x^j-y^j)} \frac{\total^d \vec{p}}{(2 \pi)^d}
\end{equation}
with the function
\begin{splitequation}
H(x,y,z,w,p) &= \omega_\vec{p} \left[ \cosh(2 \pi z) x^0 + \sinh(2 \pi z) x^1 - \cosh(2 \pi w) y^0 - \sinh(2 \pi w) y^1 \right] \\
&\quad- p^1 \left[ \sinh(2 \pi z) x^0 + \cosh(2 \pi z) x^1 - \sinh(2 \pi w) y^0 - \cosh(2 \pi w) y^1 \right] \\
&= \cosh(2 \pi z) \left( \omega_\vec{p} x^0 - p^1 x^1 \right) + \sinh(2 \pi z) \left( \omega_\vec{p} x^1 - p^1 x^0 \right) \\
&\quad- \cosh(2 \pi w) \left( \omega_\vec{p} y^0 - p^1 y^1 \right) - \sinh(2 \pi w) \left( \omega_\vec{p} y^1 - p^1 y^0 \right) \eqend{.}
\end{splitequation}
Using well-known hyperbolic function identities, we obtain the imaginary part of $H$
\begin{splitequation}
&\Im H(x,y,z,w,p) \\
&= - \sin(2 \pi b) \left[ \omega_\vec{p} \left[ \sinh(2 \pi a) x^0 + \cosh(2 \pi a) x^1 \right] - p^1 \left[ \sinh(2 \pi a) x^1 + \cosh(2 \pi a) x^0 \right] \right] \\
&\quad+ \sin(2 \pi d) \left[ \omega_\vec{p} \left[ \sinh(2 \pi c) y^0 + \cosh(2 \pi c) y^1 \right] - p^1 \left[ \sinh(2 \pi c) y^1 + \cosh(2 \pi c) y^0 \right] \right] \eqend{.}
\end{splitequation}
Let us consider the first term in brackets, for which we want to show that it is non-negative. Using hyperbolic function identities, we obtain
\begin{splitequation}
&\omega_\vec{p} \left[ \sinh(2 \pi a) x^0 + \cosh(2 \pi a) x^1 \right] - p^1 \left[ \sinh(2 \pi a) x^1 + \cosh(2 \pi a) x^0 \right] \\
&= \left( \omega_\vec{p} - \abs{p^1} \right) \left[ \cosh(2 \pi a) \left( x^1 - \abs{x^0} \right) + \mathe^{2 \pi a \sgn x^0} \abs{ x^0 } \right] \\
&\quad+ \abs{p^1} \mathe^{- 2 \pi a \sgn p^1} \left( x^1 - \abs{x^0} \right) + \abs{p^1} \left( 1 - \sgn p^1 \sgn x^0 \right) \mathe^{2 \pi a \sgn x^0} \abs{x^0} \eqend{,}
\end{splitequation}
and each term is non-negative for $x \in \mathcal{W}^r$ since then $x^1 \geq \abs{x^0}$. Analogously, the second term in brackets is seen to be non-negative, such that $\Im H(x,y,z,w,p) \leq 0$ if $\sin(2 \pi b) \geq 0$ and $\sin(2 \pi d) \leq 0$. Starting from $b = d = 0$, this thus holds in the range $\Im z = b \in [0,\frac{1}{2}]$ and $\Im w = d \in [-\frac{1}{2},0]$.

In particular, the Fourier transform of the distributional kernel $\omega_2(\Lambda_z x, \Lambda_w y)$~\eqref{eq:boosted_2pf} contains an exponentially decaying factor if $\Im H < 0$, which implies that $\omega_2\left( f_{-\frac{t}{2}}, f_\frac{t}{2} \right)$ (as a function of the real variable $t$) can be extended to a well defined bounded function of the complex variable $z$ for $z \in \mathscr{I}$, where the strip $\mathscr{I}$ is given in~\eqref{eq:complex_strip}. It furthermore follows easily that this extension is analytic in the interior of the strip and continuous on the boundaries. In conclusion, using equation~\eqref{eq:renyi_relative_fdef} we obtain for the Petz--Rényi relative entropy
\begin{splitequation}
\label{eq:petz_renyi_wedge_1}
\mathcal{S}_\alpha( W(f) \Omega \Vert \Omega ) &= \frac{1}{\alpha-1} \iint \left[ \omega_2\left( \Lambda_{-\frac{\mathi(\alpha-1)}{2}} x, \Lambda_{\frac{\mathi(\alpha-1)}{2}} y \right) - \omega_2(x,y) \right] f(x) f(y) \total^{d+1} x \total^{d+1} y \\
&= \frac{1}{\alpha-1} \lim_{\epsilon \to 0^+} \iiint \frac{1}{2 \omega_\vec{p}} \Bigl[ \mathe^{- \mathi \left[ 1 + \cos(\pi \alpha) \right] \left[ \omega_\vec{p} (x^0-y^0) - p^1 (x^1-y^1) \right]} \\
&\hspace{8em}\times \mathe^{- \sin(\pi \alpha) \left[ \omega_\vec{p} (x^1+y^1) - p^1 (x^0+y^0) \right]} - 1 \Bigr] \\
&\hspace{4em}\times f(x) f(y) \, \mathe^{\mathi \omega_\vec{p} (x^0-y^0 + \mathi \epsilon) - \mathi p^1 (x^1-y^1) - \mathi p^j (x^j-y^j)} \total^{d+1} x \total^{d+1} y \frac{\total^d \vec{p}}{(2 \pi)^d} \eqend{.}
\end{splitequation}
We can obtain another form of this result in terms of initial data associated to $f$. For this, we use that for any $f \in C_0^\infty(\mathbb{R}^{d+1})$ and any $s$ satisfying $( \partial^2 - m^2 ) s(x) = 0$, it holds that
\begin{equation}
\label{eq:initial_data}
\int s(x) f(x) \total^{d+1} x = \int \left[ s(x) \frac{\partial}{\partial x^0} (E f)(x) - (E f)(x) \frac{\partial}{\partial x^0} s(x) \right] \total^d \vec{x} \eqend{,}
\end{equation}
where $E$ is the difference between the advanced and retarded fundamental solutions of the Klein--Gordon operator $\partial^2 - m^2$, and the right-hand side is evaluated at any fixed time $x^0$. In the following, we will take $x^0 = 0$ for simplicity. Explicitly, this reads $(E f)(x) = \int E(x,y) f(y) \total^{d+1} y$ with the distributional kernel
\begin{splitequation}
\label{eq:commutator_def}
E(x,y) &= \lim_{\epsilon \to 0^+} \int \frac{\sin\left[ \omega_\vec{p} (x^0-y^0) - \vec{p} (\vec{x}-\vec{y}) \right]}{\omega_\vec{p}} \mathe^{- \epsilon \, \omega_\vec{p}} \frac{\total^d \vec{p}}{(2 \pi)^d} \\
&= - \mathi \Bigl[ \omega_2(x,y) - \omega_2(y,x) \Bigr] \eqend{,}
\end{splitequation}
and by taking suitable limits the relation~\eqref{eq:initial_data} even holds for non-smooth or non-compactly supported functions $f$ as long as both sides are finite. We note that~\eqref{eq:initial_data} is well-known and can be easily derived, see for example~\cite[App.~B]{froebmuchpapadopoulos2023}.

Because the integral kernel of the two-point function $\omega_2$~\eqref{eq:omega_2pf} is a bisolution of the Klein--Gordon equation (as can be easily checked), it follows that
\begin{splitequation}
&\iint \omega_2(x,y) f(x) f(y) \total^{d+1} x \total^{d+1} y \\
&= \iint \biggl[ \omega_2(x,y) \frac{\partial}{\partial x^0} (E f)(x) \frac{\partial}{\partial y^0} (E f)(y) - (E f)(x) \frac{\partial}{\partial x^0} \omega_2(x,y) \frac{\partial}{\partial y^0} (E f)(y) \\
&\quad- (E f)(y) \frac{\partial}{\partial y^0} \omega_2(x,y) \frac{\partial}{\partial x^0} (E f)(x) + (E f)(x) (E f)(y) \frac{\partial}{\partial x^0} \frac{\partial}{\partial y^0} \omega_2(x,y) \biggr]_{x^0 = y^0 = 0} \total^d \vec{x} \total^d \vec{y} \eqend{.}
\end{splitequation}
Also the boosted kernel~\eqref{eq:boosted_2pf} is a bisolution, such that the Petz--Rényi relative entropy~\eqref{eq:petz_renyi_wedge_1} can be written as
\begin{splitequation}
\label{eq:petz_renyi_wedge_2}
\mathcal{S}_\alpha( W(f) \Omega \Vert \Omega ) &= \frac{1}{2} \iint \biggl[ K_\alpha(x,y) \frac{\partial}{\partial x^0} (E f)(x) \frac{\partial}{\partial y^0} (E f)(y) \\
&\qquad- 2 \frac{\partial}{\partial x^0} K_\alpha(x,y) (E f)(x) \frac{\partial}{\partial y^0} (E f)(y) \\
&\qquad+ \frac{\partial}{\partial x^0} \frac{\partial}{\partial y^0} K_\alpha(x,y) (E f)(x) (E f)(y) \biggr]_{x^0 = y^0 = 0} \total^d \vec{x} \total^d \vec{y}
\end{splitequation}
with the symmetric kernel
\begin{splitequation}
K_\alpha(x,y) &= \frac{1}{\alpha-1} \biggl[ \omega_2\left( \Lambda_{-\frac{\mathi(\alpha-1)}{2}} x, \Lambda_{\frac{\mathi(\alpha-1)}{2}} y \right) + \omega_2\left( \Lambda_{-\frac{\mathi(\alpha-1)}{2}} y, \Lambda_{\frac{\mathi(\alpha-1)}{2}} x \right) \\
&\qquad\qquad- \omega_2(x,y) - \omega_2(y,x) \biggr] \eqend{.}
\end{splitequation}

In the limit $\alpha \to 1$, this expression reproduces the known result for the relative entropy of a coherent excitation in the wedge~\cite{casinigrillopontello2019,ciollilongoruzzi2020}. To see this explicitly, we note that in the limit $\alpha \to 1$ the kernel $K_\alpha$ reduces to
\begin{splitequation}
K_1(x,y) &= \lim_{\alpha \to 1^-} K_\alpha(x,y) = \lim_{\epsilon \to 0^+} \int \frac{\pi}{\omega_\vec{p}} \left[ \omega_\vec{p} (x^1+y^1) - p^1 (x^0+y^0) \right] \mathe^{- \epsilon \, \omega_\vec{p}} \\
&\hspace{10em}\times \cos\left[ \omega_\vec{p} (x^0-y^0) - p^1 (x^1-y^1) - p^j (x^j-y^j) \right] \frac{\total^d \vec{p}}{(2 \pi)^d} \\
&= \pi \left[ (x^1+y^1) \frac{\partial}{\partial x^0} + (x^0+y^0) \frac{\partial}{\partial x^1} \right] E(x,y) \eqend{.}
\end{splitequation}
where we used the relation~\eqref{eq:commutator_def} between $E$ and $\omega_2$ in the last equality. Inserting this result in~\eqref{eq:petz_renyi_wedge_2} and using that $E$ is a function of $x-y$, satisfies the relations
\begin{equation}
\label{eq:commutator_relation}
E(x,y) \bigr\rvert_{x^0 = y^0} = 0 \eqend{,} \quad \frac{\partial}{\partial x^0} E(x,y) \bigr\rvert_{x^0 = y^0} = \delta(\vec{x} - \vec{y}) \eqend{,}
\end{equation}
which follow straightforwardly from the definition~\eqref{eq:commutator_def}, and is a bisolution of the Klein--Gordon equation, we obtain
\begin{splitequation}
\label{eq:entropy_wedge}
&\mathcal{S}( W(f) \Omega \Vert \Omega ) = \lim_{\alpha \to 1^-} \mathcal{S}_\alpha( W(f) \Omega \Vert \Omega ) \\
&= \frac{\pi}{2} \iint (x^1+y^1) \biggl[ \delta(\vec{x}-\vec{y}) \frac{\partial}{\partial x^0} (E f)(x) \frac{\partial}{\partial y^0} (E f)(y) \\
&\hspace{6em}+ \left( \frac{\partial}{\partial \vec{x}} \cdot \frac{\partial}{\partial \vec{y}} + m^2 \right) \delta(\vec{x}-\vec{y}) (E f)(x) (E f)(y) \biggr]_{x^0 = y^0 = 0} \total^d \vec{x} \total^d \vec{y} \\
&= \int \pi x^1 \left[ \left[ \frac{\partial}{\partial x^0} (E f)(x) \right]^2 + \left[ \frac{\partial}{\partial \vec{x}} (E f)(x) \right]^2 + m^2 \left[ (E f)(x) \right]^2 \right]_{x^0 = 0} \total^d \vec{x} \eqend{,}
\end{splitequation}
where in the second equality we integrated by parts and used that $(E f)(x)$ vanishes as $\abs{\vec{x}} \to \infty$ for any fixed $x^0$. This is the classical Noether charge associated to the boosts, evaluated on $E f$. Recall that in the right wedge $\mathcal{W}^r$ we have $x^1 \geq 0$, such that the result is positive as required. Note that since the classical energy density for a free massive scalar field is given by $T_{00}(\phi) = \frac{1}{2} \left( \partial_{x^0} \phi \right)^2 + \frac{1}{2} \left( \partial_{\vec{x}} \phi \right)^2 + \frac{1}{2} m^2 \phi^2$, we can write the result~\eqref{eq:entropy_wedge} also in the form
\begin{equation}
\label{eq:entropy_wedge_stresstensor}
\mathcal{S}( W(f) \Omega \Vert \Omega ) = \int 2 \pi x^1 \bigl[ T_{00}\left( E f \right) \bigr]_{x^0 = 0} \total^d \vec{x} \eqend{,}
\end{equation}
which agrees with previous results for the relative entropy~\cite{casinigrillopontello2019,ciollilongoruzzi2020}.

By a similar computation we can also determine the first correction to the relative entropy, namely $\lim_{\alpha \to 1^-} \partial_\alpha \mathcal{S}_\alpha( W(f) \Omega \Vert \Omega )$. The result is lengthy and not particularly illuminating, except for the fact that it depends on the symmetric part of the two-point function $\omega_2(x,y) + \omega_2(y,x)$ and not only on the antisymmetric part $E$~\eqref{eq:commutator_def}. It is then impossible to use the relations~\eqref{eq:commutator_relation} and reduce the result to a form similar to the classical entropy of a wave packet~\eqref{eq:entropy_wedge}. Hence, the Petz--Rényi entropy is genuinely quantum.

\section{Free chiral current in a thermal state}

As a second example, we consider the chiral current $j$ on the light ray in a thermal equilibrium state, depending on the coordinate $u = t-x$ and restricted to the half-space $u \geq 0$. For this case, the modular group was determined by Borchers and Yngvason~\cite{borchersyngvason1999}. The integral kernel of the two-point function in the thermal state is given by
\begin{equation}
\omega_2(u,u') = \frac{1}{2} \lim_{\epsilon \to 0^+} \int \frac{p}{1 - \mathe^{- \beta p}} \, \mathe^{- \mathi p (u-u'-\mathi \epsilon)} \frac{\total p}{2 \pi} \eqend{,}
\end{equation}
where $\beta$ is the inverse temperature. The Fourier transform can be computed by closing the integration contour in the upper or lower half plane depending on the sign of $u-u'$ and using Cauchy's theorem, and we obtain\footnote{Our result differs from~\cite[Eq.~(4.10)]{borchersyngvason1999} by an overall factor of $- \pi/4$. This factor does not influence the results of their work, but is required for a correct normalization. In particular, since $j = \partial_u \phi$ and the two-point function of the chiral half of the massless scalar in the vacuum state reads~\cite[Eq.~(2.3b)]{schroertruong1977} $- \frac{1}{4 \pi} \ln(u-u'-\mathi \epsilon) + \text{irrelevant}$, one sees that our normalization gives the correct two-point function of the current in the zero-temperature limit $\beta \to \infty$.}
\begin{splitequation}
\label{eq:thermal_2pf}
\omega_2(u,u') &= - \frac{\pi}{4 \beta^2} \lim_{\epsilon \to 0^+} \sinh^{-2}\left[ \frac{\pi}{\beta} (u-u'-\mathi \epsilon) \right] \\
&= - \frac{1}{4 \pi} \partial_u \partial_{u'} \ln \sinh\left[ \frac{\pi}{\beta} (u-u'-\mathi \epsilon) \right] \eqend{.}
\end{splitequation}
The KMS condition reads
\begin{equation}
\label{eq:thermal_2pf_kms}
\omega_2(u-\mathi \beta,u') = \omega_2(u',u)
\end{equation}
and is easily seen to be satisfied. The action of the modular group is again geometric~\eqref{eq:geometric_modular_action} with~\cite[Eq.~(4.14)]{borchersyngvason1999}
\begin{equation}
\label{eq:chiral_current_trafo}
f_t(u) = f(L(t,u)) \eqend{,} \quad L(t,u) = \frac{\beta}{2 \pi} \ln\left[ 1 + \mathe^{2 \pi t} \left( \mathe^\frac{2 \pi u}{\beta} - 1 \right) \right] \eqend{,}
\end{equation}
where the latter function satisfies $L(0,u) = u$ and the group property $L(t, L(s,u)) = L(t+s,u)$. Restricting to $u \geq 0$, we have $L(t,u) \geq 0$ for all $t$ and thus the set of test functions with support on the positive light ray is mapped into itself.

To obtain an explicit expression for the Petz--Rényi relative entropy, we need to determine the analytic extension of the function
\begin{equation}
t \in \mathbb{R} \to \omega_2\left( f_{-\frac{t}{2}}, f_{\frac{t}{2}} \right)
\end{equation}
in the complex strip $\mathscr{I}$~\eqref{eq:complex_strip}. A transformation of coordinates together with the group property of $L$ results in
\begin{splitequation}
\omega_2\left( f_{-\frac{t}{2}}, f_{\frac{t}{2}} \right) &= \iint_0^\infty \omega_2(u,v) f\left( L\left(-\frac{t}{2}, u \right) \right) f\left( L\left(\frac{t}{2}, v \right) \right) \total u \total v \\
&= \iint_0^\infty \omega_2\left( L\left( \frac{t}{2}, u \right), L\left( - \frac{t}{2}, v \right) \right) \frac{\partial L\left( \frac{t}{2}, u \right)}{\partial u} \frac{\partial L\left( - \frac{t}{2}, v \right)}{\partial v} f(u) f(v) \total u \total v \\
&= - \frac{\pi}{4 \beta^2} \lim_{\epsilon \to 0^+} \iint_0^\infty \sinh^{-2}\left[ \frac{\pi}{\beta} \left( L\left( \frac{t}{2}, u \right) - L\left( - \frac{t}{2}, v \right) - \mathi \epsilon \right) \right] \\
&\qquad\qquad\times \mathe^{\frac{2 \pi}{\beta} (u+v)} \mathe^{- \frac{2 \pi}{\beta} L\left( \frac{t}{2}, u \right)} \mathe^{- \frac{2 \pi}{\beta} L\left( - \frac{t}{2}, v \right)} f(u) f(v) \total u \total v \\
&= - \frac{\pi}{\beta^2} \lim_{\epsilon \to 0^+} \iint_0^\infty \left[ \frac{\mathe^{\frac{\pi}{\beta} (u+v)}}{\left[ 1 + \mathe^{\pi t} \left( \mathe^\frac{2 \pi u}{\beta} - 1 \right) \right] \mathe^{- \mathi \frac{\pi}{\beta} \epsilon} - \left[ 1 + \mathe^{- \pi t} \left( \mathe^\frac{2 \pi v}{\beta} - 1 \right) \right] \mathe^{\mathi \frac{\pi}{\beta} \epsilon}} \right]^2 \\
&\qquad\qquad\times f(u) f(v) \total u \total v \eqend{.}
\end{splitequation}
In the limit $\epsilon \to 0^+$, only the sign of the coefficient in front of $\epsilon$ matters, such that we can replace (for real $t$) the denominator by $\mathe^{\pi t} \left( \mathe^\frac{2 \pi u}{\beta} - 1 \right) - \mathe^{- \pi t} \left( \mathe^\frac{2 \pi v}{\beta} - 1 \right) - \mathi \epsilon$. Moreover, we can extract derivatives with respect to $u$ and $v$ and integrate by parts, such that
\begin{splitequation}
\omega_2\left( f_{-\frac{t}{2}}, f_{\frac{t}{2}} \right) &= - \frac{1}{4 \pi} \lim_{\epsilon \to 0^+} \iint_0^\infty \ln\left[ \mathe^{\pi t} \left( \mathe^\frac{2 \pi u}{\beta} - 1 \right) - \mathe^{- \pi t} \left( \mathe^\frac{2 \pi v}{\beta} - 1 \right) - \mathi \epsilon \right] \\
&\qquad\qquad\times f'(u) f'(v) \total u \total v \eqend{.}
\end{splitequation}

The distributional kernel can now be extended to an analytic function in the strip $\mathscr{I}$ in the obvious way. Replacing $t \to z = t - \mathi r$, we obtain
\begin{splitequation}
&- \frac{1}{4 \pi} \lim_{\epsilon \to 0^+} \iint_0^\infty \ln\bigg[ \cos(\pi r) \left[ \mathe^{\pi t} \left( \mathe^\frac{2 \pi u}{\beta} - 1 \right) - \mathe^{- \pi t} \left( \mathe^\frac{2 \pi v}{\beta} - 1 \right) \right] \\
&\hspace{8em}- \mathi \sin(\pi r) \left[ \mathe^{\pi t} \left( \mathe^\frac{2 \pi u}{\beta} - 1 \right) + \mathe^{- \pi t} \left( \mathe^\frac{2 \pi v}{\beta} - 1 \right) \right] - \mathi \epsilon \bigg] f'(u) f'(v) \total u \total v \eqend{,}
\end{splitequation}
and for $r \in (0,1)$ the argument of the logarithm is always non-zero such that we can take the limit $\epsilon \to 0^+$ inside the integral. Moreover, since $u,v \geq 0$ one verifies by a long but straightforward case-by-case analysis that the complex argument of the argument of the logarithm is in the range $(-\pi,0)$ where the logarithm is an analytic function. Using equation~\eqref{eq:renyi_relative_fdef} we thus obtain for the Petz--Rényi relative entropy
\begin{splitequation}
\label{eq:petz_renyi_lightray}
\mathcal{S}_\alpha( W(f) \Omega \Vert \Omega ) &= - \frac{1}{4 \pi (\alpha-1)} \lim_{\epsilon \to 0^+} \iint_0^\infty \\
&\qquad\times \Biggl[ \ln\left[ - \cos(\pi \alpha) \left( \mathe^\frac{2 \pi u}{\beta} - \mathe^\frac{2 \pi v}{\beta} \right) - \mathi \sin(\pi \alpha) \left( \mathe^\frac{2 \pi u}{\beta} + \mathe^\frac{2 \pi v}{\beta} - 2 \right) - \mathi \epsilon \right] \\
&\qquad\qquad- \ln\left( \mathe^\frac{2 \pi u}{\beta} - \mathe^\frac{2 \pi v}{\beta} - \mathi \epsilon \right) \Biggr] f'(u) f'(v) \total u \total v \eqend{.}
\end{splitequation}
Integrating the $u$ and $v$ derivatives by parts, we also obtain the alternative expression
\begin{splitequation}
\label{eq:petz_renyi_lightray_alt}
\mathcal{S}_\alpha( W(f) \Omega \Vert \Omega ) &= - \frac{\pi^2}{\beta^2} \frac{1}{4 \pi (\alpha-1)} \lim_{\epsilon \to 0^+} \iint_0^\infty \\
&\qquad\times \Biggl[ \left[ \sinh\left[ \frac{\pi}{\beta} ( u-v-\mathi \epsilon) + \mathi \pi \alpha \right] - \mathi \sin(\pi \alpha) \mathe^{- \frac{\pi (u+v)}{\beta}} \right]^{-2} \\
&\qquad\qquad- \sinh^{-2}\left[ \frac{\pi}{\beta} (u-v-\mathi \epsilon) \right] \Biggr] f(u) f(v) \total u \total v \eqend{.}
\end{splitequation}

To determine the limit of the Petz--Rényi relative entropy~\eqref{eq:petz_renyi_lightray} as $\alpha \to 1$, we use l'H{\^o}pital's rule and obtain
\begin{splitequation}
\label{eq:relative_entropy_lightray}
&\mathcal{S}( W(f) \Omega \Vert \Omega ) = \lim_{\alpha \to 1^-} \mathcal{S}_\alpha( W(f) \Omega \Vert \Omega ) \\
&\quad= - \frac{\mathi}{4} \lim_{\epsilon \to 0^+} \iint_0^\infty \frac{\mathe^\frac{2 \pi u}{\beta} + \mathe^\frac{2 \pi v}{\beta} - 2}{\mathe^\frac{2 \pi u}{\beta} - \mathe^\frac{2 \pi v}{\beta} - \mathi \epsilon} f'(u) f'(v) \total u \total v \\
&\quad= - \frac{1}{4} \lim_{\epsilon \to 0^+} \iint_0^\infty \frac{\epsilon \left( \mathe^\frac{2 \pi u}{\beta} + \mathe^\frac{2 \pi v}{\beta} - 2 \right)}{\left( \mathe^\frac{2 \pi u}{\beta} - \mathe^\frac{2 \pi v}{\beta} - \mathi \epsilon \right) \left( \mathe^\frac{2 \pi v}{\beta} - \mathe^\frac{2 \pi u}{\beta} - \mathi \epsilon \right)} f'(u) f'(v) \total u \total v \\
&\quad= \frac{\pi}{4} \iint_0^\infty \left( \mathe^\frac{2 \pi u}{\beta} + \mathe^\frac{2 \pi v}{\beta} - 2 \right) \delta\left( \mathe^\frac{2 \pi u}{\beta} - \mathe^\frac{2 \pi v}{\beta} \right) f'(u) f'(v) \total u \total v \\
&\quad= \frac{\beta}{4} \int_0^\infty \left( 1 - \mathe^{- \frac{2 \pi u}{\beta}} \right) \bigl[ f'(u) \bigr]^2 \total u \eqend{,}
\end{splitequation}
where in the second equality we symmetrized the integrand in $u$ and $v$. Since for the chiral current the classical energy density is given by $T_{00}(j) = \frac{1}{2} (j')^2$, we can write this also as
\begin{equation}
\label{eq:relative_entropy_lightray_stresstensor}
\mathcal{S}( W(f) \Omega \Vert \Omega ) = \frac{\beta}{2} \int_0^\infty \left( 1 - \mathe^\frac{- 2 \pi u}{\beta} \right) T_{00}(f) \total u \eqend{.}
\end{equation}
This agrees with known results for the relative entropy~\cite{bostelmanncadamurodelvecchio2022,blancocasinilestonrosso2018}, up to the overall normalization.

Moreover, we can compute the first corrections to the relative entropy for $\alpha$ close to $1$. Namely, we obtain
\begin{splitequation}
\lim_{\alpha \to 1^-} \partial_\alpha \mathcal{S}_\alpha( W(f) \Omega \Vert \Omega ) &= - \frac{\pi}{8} \lim_{\epsilon \to 0^+} \iint_0^\infty f'(u) f'(v) \\
&\qquad\times \frac{4 \left( \mathe^\frac{2 \pi u}{\beta} - 1 \right) \left( \mathe^\frac{2 \pi v}{\beta} - 1 \right) + \mathi \epsilon \left( \mathe^\frac{2 \pi u}{\beta} - \mathe^\frac{2 \pi v}{\beta} \right)}{\left( \mathe^\frac{2 \pi u}{\beta} - \mathe^\frac{2 \pi v}{\beta} - \mathi \epsilon \right)^2} \total u \total v \eqend{,}
\end{splitequation}
and since by our general results $\mathcal{S}_\alpha$ is monotonically increasing in $\alpha$ this must be non-negative. We see this by rewriting the fraction in the integrand as
\begin{equation}
\frac{\left( 1 - \mathe^{- \frac{2 \pi u}{\beta}} \right) \left( 1 - \mathe^{- \frac{2 \pi v}{\beta}} \right)}{\left[ \sinh\left[ \frac{\pi}{\beta} (u-v) \right] - \mathi \epsilon \right]^2} + \frac{\mathi \epsilon \sinh\left[ \frac{\pi}{\beta} (u-v) \right]}{\left[ \sinh\left[ \frac{\pi}{\beta} (u-v) \right] - \mathi \epsilon \right]^2} \eqend{,}
\end{equation}
taking into account that in the limit $\epsilon \to 0^+$ only the sign of the coefficient in front of $\epsilon$ matters, such that we can rescale $\epsilon$ by any positive function. In the limit $\epsilon \to 0^+$, the second term vanishes as a distribution, while the first is nothing else but the integral kernel of the two-point function~\eqref{eq:thermal_2pf} up to a rescaling. It thus follows that
\begin{equation}
\label{eq:petz_renyi_lightray_corr}
\lim_{\alpha \to 1^-} \partial_\alpha \mathcal{S}_\alpha( W(f) \Omega \Vert \Omega ) = \frac{\beta^2}{2} \iint_0^\infty f'(u) f'(v) \left( 1 - \mathe^{- \frac{2 \pi u}{\beta}} \right) \left( 1 - \mathe^{- \frac{2 \pi v}{\beta}} \right) \, \omega_2(u,v) \total u \total v \geq 0 \eqend{,}
\end{equation}
which is positive because $\omega_2$ is a positive-definite kernel. We note that as for the scalar field in the wedge, this depends also on the symmetric part of the two-point function, and hence is genuinely quantum.

Lastly, we consider the temperature dependence of the Petz--Rényi relative entropy. For generic $\alpha$, the integral kernel of $\mathcal{S}_\alpha( W(f) \Omega \Vert \Omega )$~\eqref{eq:petz_renyi_lightray} depends on $\beta$ in a complicated way, and we thus consider only the relative entropy~\eqref{eq:relative_entropy_lightray}. We obtain straightforwardly that
\begin{equation}
\label{eq:beta_derivative_relative_entropy}
\partial_\beta \mathcal{S}( W(f) \Omega \Vert \Omega ) = \frac{1}{4} \int_0^\infty \left( 1 - \mathe^{- \frac{2 \pi u}{\beta}} - \frac{2 \pi u}{\beta} \mathe^{- \frac{2 \pi u}{\beta}} \right) \bigl[ f'(u) \bigr]^2 \total u \geq 0 \eqend{,}
\end{equation}
which is positive since $1 - \mathe^{-x} - x \mathe^{-x} \geq 0$ for $x \geq 0$.\footnote{This follows because the left-hand side vanishes for $x = 0$ and its derivative is positive for all $x \geq 0$.} In particular, for a fixed coherent excitation of the thermal state (known as thermal coherent state) implemented by the Weyl operator $W(f)$ with $f$ independent of temperature, the relative entropy is maximal for $\beta \to \infty$, corresponding to zero temperature. In this limit, we obtain
\begin{equation}\label{eq:beta_infinity_limit}
\lim_{\beta \to \infty} \mathcal{S}( W(f) \Omega \Vert \Omega ) = \frac{1}{4} \int_0^\infty 2 \pi u \bigl[ f'(u) \bigr]^2 \total u = \int_0^\infty \pi u T_{00}(f) \total u \eqend{,}
\end{equation}
where $T_{00}(j) = \frac{1}{2} (j')^2$ is the classical energy density of the chiral current. Apart from the overall normalization (due to the fact that we consider a light ray and not an equal-time Cauchy surface), this agrees with the expression~\eqref{eq:entropy_wedge_stresstensor} for the free scalar field in the wedge. This is in agreement with the fact that in the limit $\beta \to \infty$ the thermal state reduces to the usual vacuum state and the modular Hamiltonian for the wedge to the generator of boosts.

On the other hand, in the infinite-temperature limit $\beta \to 0$ we obtain
\begin{equation}\label{eq:beta_zero_limit_relative}
\lim_{\beta \to 0} \mathcal{S}( W(f) \Omega \Vert \Omega ) = 0 \eqend{,}
\end{equation}
and in fact for the full Petz--Rényi relative entropy~\eqref{eq:petz_renyi_lightray_alt} for any $\alpha \in [0,1)$
\begin{equation}\label{eq:beta_zero_limit_Rènyi}
\lim_{\beta \to 0} \mathcal{S}_\alpha( W(f) \Omega \Vert \Omega ) = 0 \eqend{.}
\end{equation}
This result is physically explained by the fact that in the limit of infinite temperature the coherent excitation is immediately dissipated, making the states $\ket{\Omega}$ and $W(f) \ket{\Omega}$ indistinguishable. From a mathematical point of view, the state $\omega^\beta$ becomes a tracial state in this limit, which can be easily inferred by the KMS condition~\eqref{eq:thermal_2pf_kms} satisfied with respect to the usual time evolution in the limit $\beta \to 0$. In fact, we see that the integral kernel of the two-point function~\eqref{eq:thermal_2pf} reduces to
\begin{splitequation}
\omega_2(u,u') \approx - \frac{1}{4 \beta} \partial_u \partial_{u'} \abs{u-u'} = \frac{1}{2 \beta} \delta(u-u')
\end{splitequation}
and becomes ultralocal. We see that the symmetric and antisymmetric parts of the two-point function are of different order in $\beta$, and in particular the antisymmetric part is subleading as $\beta \to 0$, which immediately implies the tracial property for the state $\omega^\beta$.

The tracial property is of course also satisfied by $\omega^\beta$ once we restrict it to the von Neumann subalgebra generated by $W(f)$ with $\supp f \subset \mathbb{R}^+$, namely the Weyl operators supported on the right wedge. Modular theory for tracial states turns out to be particularly simple, since for these the modular operator coincides with the identity~\cite[Sec.~2.5]{brattelirobinson1}. Indeed, given a faithful tracial state $\tau$ on a von Neumann algebra $\mathfrak{A}$ implemented by a cyclic and separating vector $\ket{\Psi}$ it holds that
\begin{equation}
\norm{ S_\Psi a \ket{\Psi} }^2 = \norm{ a^\dagger \ket{\Psi} }^2 = \bra{\Psi} a a^\dagger \ket{\Psi} = \tau(a a^\dagger) = \tau(a^\dagger a) = \bra{\Psi} a^\dagger a \ket{\Psi} = \norm{ a \ket{\Psi} }^2 \eqend{.}
\end{equation}
That is, for faithful tracial states $S_\Psi$ is a unitary operator, which by the uniqueness of the polar decomposition implies that $\Delta_\Psi = \1$. We can directly check this property for the chiral current in the thermal state by taking the limit $\beta \to 0$ of the transformation $L(t,u)$~\eqref{eq:chiral_current_trafo}:
\begin{equation}
\lim_{\beta \to 0} L(t,u) = u \eqend{.}
\end{equation}
It follows that the action~\eqref{eq:geometric_modular_action} of the modular group is the identity for all $t$, which is only possible for $\Delta_\Omega = \1$. This then also implies that the relative entropy, as well as the Petz--Rényi relative entropy, vanish in this limit.

\section{Outlook}

We have studied the Petz--Rényi relative entropy $\mathcal{S}_\alpha(\Psi \vert \Phi)$ between two states $\ket{\Psi}$ and $\ket{\Phi}$ of a von Neumann algebra $\mathfrak{A}$ using modular theory. We have shown that it is positive, monotonous, and bounded from above by the Araki--Uhlmann relative entropy $\mathcal{S}(\Psi \vert \Phi)$. We have further shown that for unitary excitations of a common state $\ket{\Omega}$, it can be computed using only the modular Hamiltonian of $\ket{\Omega}$, and that the computation can be done by analytically continuing the modular flow generated by this Hamiltonian. For free bosonic QFTs and under the assumption of geometric modular action~\eqref{eq:geometric_modular_action}, we have then shown that the $\mathcal{S}_\alpha(\Psi \vert \Phi)$ can be obtained using only the two-point function $\omega_2$ of $\ket{\Omega}$~\eqref{eq:renyi_relative_unitary}. In contrast to the Araki--Uhlmann relative entropy, apart from the antisymmetric part of the two-point function (the sympletic form), here also the symmetric part contributes in general. In particular, this entails that the Petz--Rényi relative entropy can not be expressed using just the classical entropy of a wave packet, such that it contains genuine quantum effects.

As a first example, we have computed $\mathcal{S}_\alpha(\Psi \vert \Phi)$ for a free massive scalar field in the Minkowski vacuum state, with $\mathfrak{A}$ being the algebra of Weyl operators in the right wedge. We have shown explicitly how the Petz--Rényi entropy~\eqref{eq:petz_renyi_wedge_2} is computed by the analytic continuation of the modular flow, and we have verified that in the limit $\alpha \to 1$ it reduces to the known results~\eqref{eq:entropy_wedge}, \eqref{eq:entropy_wedge_stresstensor} for the relative entropy~\cite{casinigrillopontello2019,ciollilongoruzzi2020}. In the future, it would be very interesting to extend our work also to fermions, where results for the relative entropy are also available~\cite{galandamuchverch2023}.

As a second example we computed $\mathcal{S}_\alpha(\Psi \vert \Phi)$ for the free chiral current $j$ in a thermal equilibrium (KMS) state on Minkowski spacetime. In this case, $\mathfrak{A}$ is the algebra generated by Weyl operators supported on a light ray and restricted to the half-space $u = t-x \geq 0$. We have employed the results of~\cite{borchersyngvason1999} for the corresponding modular group to derive an explicit expression for the Petz--Rényi entropy~\eqref{eq:petz_renyi_lightray}. We have also verified that in the limit $\alpha \to 1$ it reduces to the known results~\eqref{eq:relative_entropy_lightray}, \eqref{eq:relative_entropy_lightray_stresstensor} for the relative entropy~\cite{bostelmanncadamurodelvecchio2022,blancocasinilestonrosso2018} and that the first correction (given by the derivative with respect to $\alpha$ in the limit $\alpha \to 1$) is positive, as required on general grounds~\cite{bertascholztomamichel2018}. Finally, we have studied the dependence of the relative entropy on the inverse temperature $\beta$~\eqref{eq:beta_derivative_relative_entropy}, showing that it is a monotonically increasing function. In particular, in the limit $\beta \to \infty$ (vanishing temperature) we have recovered the expected result~\eqref{eq:beta_infinity_limit}, in agreement with the fact that the KMS state reduces to the usual Minkowski vacuum state in this limit. On the contrary, we have shown that in the limit $\beta \to 0$ the relative entropy vanishes~\eqref{eq:beta_zero_limit_relative} (and in fact the full Petz--Rényi relative entropy vanishes as well~\eqref{eq:beta_zero_limit_Rènyi}). We have shown that this result is a consequence of the tracial property of the thermal equilibrium state in the limit of infinite temperature. Also in this case, it would be very interesting to extend our work to fermions.

\begin{acknowledgments}
This work has been funded by the Deutsche Forschungsgemeinschaft (DFG, German Research Foundation) --- project no. 396692871 within the Emmy Noether grant CA1850/1-1.
\end{acknowledgments}

\appendix

\subsection*{Conflict of interest statement}

On behalf of all authors, the corresponding author states that there is no conflict of interest.

\subsection*{Data availability statement}

This manuscript has no associated data.

\bibliography{literature}

\end{document}